\documentclass[12pt,preprint]{aastex}
\usepackage{amsmath}
\usepackage{color}
\usepackage{graphicx}
\newcommand{\eqb}{\begin{equation}}
\newcommand{\eqe}{\end{equation}}
\begin{document}

\title{Induced scattering limits on fast radio bursts from stellar coronae}
\author{Yuri Lyubarsky$^1$, Sofiya Ostrovska$^2$}
 \affil{$^1$Physics Department, Ben-Gurion University, P.O.B. 653, Beer-Sheva 84105, Israel\\
 $^2$Department of Mathematics, Atilim University, Incek 06836, Ankara, Turkey}
\begin{abstract}
The origin of fast radio bursts remains a puzzle. Suggestions have been made that they are produced within the Earth atmosphere, in stellar coronae, in other galaxies or at cosmological distances.  If they are extraterrestrial, the implied brightness temperature is very high, and therefore, the induced scattering places constraints on possible models. In  this paper, constraints are obtained on flares from coronae of nearby stars.
It is shown that the radio pulses with the observed power could not be generated if the plasma density within and in the nearest vicinity of the source is as high as it is necessary in order to provide the observed dispersion measure. However, one cannot exclude a possibility that the pulses are generated within a bubble with a very low density and pass through the dense plasma only in the outer corona. 
\end{abstract}
\maketitle

\section{Introduction}
Since the first fast radio burst (FRB) was discovered \citep{Lorimer07}, nearly a dozen such events has been reported \citep{Keane11,Thornton13,Burke14,Spitler14,Petroff15,Ravi_etal15}. Inasmuch as these millisecond flares exhibit very large dispersion measure, significantly exceeding the Galactic one (see, however, \citet{Bannister_Madsen14}),  the currently favored interpretation is that they occur at cosmological distances. The implied tremendous  energy release (isotropic equivalent $\sim 10^{40}$ erg) in the radio band should be attributed to exotic events; among the discussed are magnetar hyper flares \citep{PopovPostnov07,PostnovPopov13,Thornton13,Katz13,Lyubarsky14}, supernova explosion in a binary containing a neutron star \citep{EgorovPostnov09}, collapse of a supermassive neutron star \citep{FalsckeRezzolla13,Zhang14}, binary white dwarf or neutron star merger \citep{Keane12,Kashiyama13,Lipunov13,Totani13,Ravi_Lasky14}, evaporation of primordial black hole \citep{Keane12}, an asteroid/comet impact with a neutron star \citep{Geng_Huang15}, supergiant pulses from pulsars \citep{Connor_etal15,Katz15}, quark nova \citep{Shand15}. The possible terrestrial origin is discussed by \citet{Burke11} and \citet{Kulkarni14}.

\citet{Loeb14} suggested that the sources of FRBs are flaring stars in our Galaxy. In this case, large dispersion measure is due to the stellar corona with the density $10^{8-9}$ cm$^{-3}$ extended to the distance $\sim 10^{12-13}$ cm. \citet{Maoz15} added credibility to the model finding flare stars in FRB fields and also showing that the previous concerns about too high free-free absorption in the corona \citep{Luan14} or possible significant deviations from the $f^{-2}$ dependence of the pulse arrival times \citep{Dennison14,Tuntsov14} may be alleviated. Even though the required energy release is not extraordinary high in this model, the inferred brightness temperature is still very high so that some coherent radiation mechanism is assumed. The authors mention the cyclotron instability as a candidate. But independently of the radiation mechanism, the induced scattering on the relatively dense plasma in the corona should unavoidably affect  the outgoing radiation.  It will be demonstrated in this paper that FRBs could be generated in coronae of nearby stars only if the density within and in the nearest vicinity of the source does not exceed $\sim 10^5$ cm$^{-3}$, much smaller than it is necessary in order to explain the observed high dispersion measure. The only option available for having a FRB from a star in our Galaxy is to assume that the flare occurs in a low density bubble embedded into a high density corona.

\section{Induced Compton scattering in the vicinity of the source}

The kinetic equation for the induced Compton scattering in the
non-relativistic plasma is written as (e.g., \citet{Wilson82}){\color{red}:}
 \eqb
\frac{dn(\nu,\mathbf{\Omega})
}{dt}=
\frac{3\sigma_T}{8\pi}N \frac{h}{m_ec}n(\nu,\mathbf{\Omega})
 \int(\mathbf{e\cdot e}_1)^2
(1-\mathbf{\Omega\cdot\Omega}_1)\frac{\partial
\nu^2n(\nu,\mathbf{\Omega}_1)}{\partial\nu}d\mathbf{\Omega}_1;
  \label{kinComp}\eqe
where
$n(\nu,\mathbf{\Omega})$ is the photon occupation number in the direction $\mathbf{\Omega}$, $N$ the electron number
density, $\mathbf{e}$ the polarization vector, and
\eqb
\frac{d
}{dt}=\frac{\partial }{\partial t}+
c(\mathbf{\Omega\cdot\nabla})
\eqe
the derivative along the ray.

The induced scattering does not affect the escape time of photons from the source but redistributes photons towards lower frequencies where they are ultimately absorbed \citep{Sunyaev71}. The kinetic equation for the induced scattering may be presented in the form
 \eqb
\frac{dn}{dt}= A\{n\}n.
 \label{dn/dt}\eqe
The process is efficient if the frequency redistribution rate, $A\{n\}$, exceeds the escape rate (for optically thin sources, the last is just the light travel time, $r_0/c$). 

Within and in the nearest vicinity of the source, the radiation subtends a large angle; therefore, for rough estimates, one can neglect the angular dependence of $n$ in the rhs of  Eq. (\ref{kinComp}), which yields
\eqb
\frac{dn}{dt}=\sigma_TN\frac{\hbar n}{m_ec}\frac{\partial
\nu^2n}{\partial\nu}
\label{Kompaneets}\eqe
In the spatially homogeneous case, this equation is directly obtained from the Kompaneets equation by passing to the limit as the radiation brightness temperature significantly exceeds the electron temperature (e.g., \citet{Sunyaev71}).
It follows immediately from  Eq.  (\ref{Kompaneets}) that the radiation with the brightness temperature $T_b\equiv\hbar\nu n/k_B$  is unable to escape from a source with the Thomson depth $\tau_T$ if \citep{Sunyaev71,Wilson82,Coppi_etal93}
\eqb
\tau_{\rm ind}\equiv\frac{k_BT_b}{m_ec^2}\tau_T>1.
\label{tau_ind}\eqe
The effective optical depth, $\tau_{\rm ind}$, is the ratio of the escape time to the frequency redistribution time therefore at $\tau_{\rm ind}\gg 1$, only a fraction $1/\tau_{\rm ind}$ of photons escapes; the rest is redistributed towards smaller frequencies. The total number of photons is conserved in the scattering process therefore the photon phase density and the brightness temperature increase when the radiation is redistributed towards smaller frequencies so that the rate of redistribution increases further out. If no competition process came into play, the photon Bose condensation would occur at zero frequency \citep{Zeldovich_Levich68,Zeldovich_etal72}. In reality, the photons are eventually absorbed because, e.g., when the frequency approaches the plasma frequency, the free-free absorption coefficient goes to infinity.

Typical parameters of the FRB are: the pulse duration, $\Delta t\sim 1$ ms, the wavelength, $\lambda\approx 20$ cm, and the flux, $S\sim 1$ Jy. The size of the source is limited by the sound travel time; for a non-relativistic source like stellar corona one gets
\eqb
r_0\le\Delta t v_s=3\times 10^7 \frac{v_s}c\Delta t_{-3}\,\rm cm,
\eqe
where $\Delta t_{-3}=\Delta t/1$ ms, and,  $v_s$ is the sound velocity in the source. In some cases, the sound velocity should be substituted by the Alfven velocity, although    
 in any event, these velocities are below the speed of light.
One can now estimate the brightness temperature in the burst as:
\eqb
T_b=\frac{\lambda^2 SD^2}{2\pi r^2_0 k_B}>5\times 10^{21}\lambda^2_{20}S_{\rm Jy}\left(\frac{D_{300}c}{\Delta t_{-3}v_s}\right)^2\,\rm K,
\eqe
where $\lambda_{20}=\lambda/20$ cm, $S_{\rm Jy}=S/1$ Jy, and $D_{300}=D/300$ pc.
If the density in the corona is $N=10^9N_9$ cm$^{-3}$, the effective optical depth of the FRB source for the induced Compton scattering is estimated as
\eqb
\tau_{\rm ind}=\frac{k_BT_b}{m_ec^2}\sigma_TNr_0>1.6\times 10^4\frac{\lambda_{20}^2S_{\rm Jy}D_{300}^2N_9}{\Delta t_{-3}}\frac c{v_s}.
\eqe
It can be noticed that the radiation with the brightness temperature observed in FRBs could not escape from the source if the plasma density is as high as it is necessary in order to provide the observed dispersion measure.

\section{Induced Compton scattering in outer corona}

In the previous section, we considered the induced scattering within and in the nearest vicinity of the source, and  thus the condition (\ref{tau_ind}) was used, which assumes that the radiation subtends a large solid angle. The conclusion was reached that the radiation with the required brightness temperature could not escape if the plasma density is the same as in the corona. However, 
 one can speculate that 
 the FRB is produced within a bubble with the density much smaller than that in the main body of the corona.
  Alternatively, one can assume that the magnetic field in the vicinity of the source is so high that the Larmor frequency significantly esceeds the radiation frequency; then the scattering is  suppressed for extraordinary mode (in which the rotation of polarization vector is opposite to the electron Larmor rotation). In both of these cases, one can find parameters such that the effective optical depth for the induced scattering within and around the source remains less than unity. However, one has still to assume that the emitted radiation passes through a dense and weakly magnetized outer corona in order to obtain the observed $f^{-2}$ dependence of the pulse arrival time. In this section, we consider the induced scattering in the corona far from the source.

At any point far from the source, 
the radiation field subtends a small solid angle forming a narrow local radiation beam.  The induced
scattering rate is proportional to the number of photons already
available in the final state,  thence,   the scattering initially
occurs within the primary beam where the radiation
density is high. However, when the primary radiation is highly directed,
 the recoil
factor $1-\mathbf{\Omega\cdot\Omega}_1$ in the rhs of Eq. (\ref{kinComp}) makes the scattering
within the beam inefficient. In this case, the scattering outside the primary radiation beam
dominates \citep{Coppi_etal93} because according to Eq.(\ref{dn/dt}), even weak
background radiation (created, e.g., by spontaneous
scattering) grows exponentially in the course of the induced scattering so that the energy of the
scattered radiation becomes eventually comparable with the energy
density in the primary beam.

The effect of the induced scattering on short, bright radio pulses passing a plasma screen at large distances from the source was studied by \citet{Lyubarsky08}. He introduced the effective optical depth as
 \eqb
\tau_{\rm ind, pulse}=\frac{3\sigma_T}{8\pi} \frac{\lambda^2NS}{m_ec}\left(\frac{D}{r}\right)^2Z,
 \label{tau1}\eqe
 where the factor $Z$ is determined by the pulse duration and shape, $r$ the distance from the source to the scattering screen. In the simplest case of  a rectangular pulse, $Z=\Delta t$. 
This expression assumes that the width of the screen exceeds  the pulse width, $c\Delta t$; then the effective optical depth is independent of the screen width. It is shown that the induced scattering does not affect the pulse if
 \eqb
 \tau_{\rm ind, pulse}\la 10.
 \label{transp}\eqe
The factor of ten arises due to the fact that, in this situation, the radiation at large angles to the propagation direction of the primary radiation should have enough time to grow from a very low background level.

Substituting typical parameters of FRB, one can write the transparency condition (\ref{transp}) as
 \eqb
 r>10^9\lambda_{20}D_{300}N_9^{1/2}S_{\rm Jy}^{1/2}\Delta t_{-3}^{1/2}\,\rm cm.
 \eqe
One sees that the induced scattering in the outer corona could not affect  propagation of the pulse.
\section{Raman scattering in outer corona}

The propagation of a high brightness temperature pulse may be also affected by induced Raman scattering. In this process, the photon decays into another photon and a Langmuir plasmon. In other words,  the radio emission is scattered off Lanmuir plasmons,  which it generates.  The scattering rate depends on the intensity of Langmuir turbulence, which is limited by the Landau damping and collisional decay of plasmons. One can neglect the plasmon decay if the radiation power is high enough; in terms of the observed parameters the critical observed flux, above which one can neglect the plasmon decay, is found as \citep{Thompson_etal94,Lyubarsky08}
\eqb
S_{\kappa}=\frac{16\pi m_e\nu\nu_p}{3\sigma_TcN}\left(\frac rD\right)^2\kappa,
\eqe
where $\nu_p$ is the plasma frequency, $\kappa$ the plasmon decay rate. If the last is determined by the electron-ion collisions, one finds
\eqb
S_{\kappa}=1.3\times 10^{-3}\frac{N^{1/2}_9}{\lambda_{20}T_6^{3/2}}\left(\frac{r_9}{D_{300}}\right)^2\,\rm Jy,
\eqe
where $T=10^6T_6$ K is the plasma temperature.

In order to get the maximal scattering rate, let us neglect plasmon decay.
In this case, the effective optical depth with respect to the induced Raman scattering for a pulse with the duration $\Delta t$ is found as \citep{Lyubarsky08}
\eqb
\tau_{R}=\frac{3\sigma_T cNS}{m_e\nu\nu_p}\left(\frac Dr\right)^2\Delta t.
\eqe
The transparency condition looks, like in the case of the induced Compton scattering, see Eq. (\ref{transp}), as $\tau_R\la 10$. This yields the condition for the minimal distance to the high density region
\eqb
r>2.2\times 10^9 N_9^{1/4} \left(S_{\rm Jy}\Delta t_{-3}\right)^{1/2}D_{300}\,\rm cm.
\eqe
 One sees that the induced Raman scattering does not place much more severe restriction on the propagation in the outer corona than the induced Compton scattering.

\section{Conclusions}

In this paper, we studied the induced scattering of short radio pulses (having in mind FRBs) in stellar coronae. The rate of the induced scattering is highest within and in the vicinity of the source, where the radiation energy density is maximal. Our estimate show that a FRB could be emitted only if the density in and around the source is as small as
\eqb
N\la 10^5 \frac{\Delta t_{-3}}{\lambda_{20}^2S_{\rm Jy}D^2_{300}}\frac{v_s}c\, {\rm cm}^{-3}
\eqe
This is well below the density $N\sim 10^{8-9}$ cm$^{-3}$ necessary in order to provide the dispersion measure observed in FRBs.

Of course the dispersion measure is acquired at distances $\sim 10^{12-13}$ cm, much larger than the source size, so that one can not exclude that FRBs are generated within very low density bubbles, which are formed by some reason in the dense corona. We estimated the induced scattering (both Compton and Raman) at large distances from the source and found that pulses with the parameters of FRBs could propagate through the corona if they meet the dense plasma at distances not less than $\sim 10^9$ cm from the source. Therefore our final conclusion is that  FRBs could be produced in stellar coronae but only if a rather non-trivial density distribution is maintained at least during the burst.


\acknowledgements
We are grateful to an anonymous referee for constructive criticism. YL acknowledges support from the Israeli Science Foundation under the grant 719/14.

\end{document}